\documentclass[final,1p,times]{elsarticle} 

\usepackage{lineno,hyperref}
\usepackage{color}

\journal{Elsevier}
\begin{document}

\begin{frontmatter}

\title{Computational study of alpha-ray induced electron excitation in diamonds for radiation detection}

\author[address1]{Atsuhiro Umemoto\corref{mycorrespondingauthor}}
\address[address1]{International Center for Quantum-field Measurement Systems for Studies of the Universe and Particles (QUP),
High Energy Accelerator Research Organization (KEK),\\ 1-1 Oho, Tsukuba, Ibaraki 305-0801,Japan}
\cortext[mycorrespondingauthor]{Corresponding author}
%\ead{aumemoto@post.kek.jp}

\author[address2]{Yoshiyuki Miyamoto}

\address[address2]{The National Institute of Advanced Industrial Science and Technology (AIST), 1-1-1 Umezono, Tsukuba, Ibaraki, 305-8568, Japan}

\begin{abstract}
To investigate the mechanism of radiation detection in diamonds, we developed a real-time time-dependent density functional theory–based calculation scheme to evaluate changes in the density of states induced by alpha-ray irradiation. A bulk diamond structural model was constructed, with impurities optionally introduced to assess their effect on electronic excitation. Simulations revealed that the passage of high-speed helium ions, representing alpha particles, produced significant electronic excitation in the diamond model.
Subsequent calculations of the excited-state dynamics after ion removal indicated that excitation can persist for several hundred femtoseconds without triggering nonradiative relaxation. These findings demonstrate that the proposed approach offers a robust theoretical framework for evaluating the performance of diamond-based radiation detectors.

\end{abstract}


\end{frontmatter}

%\linenumbers

\section{Introduction}
The application of diamonds as radiation detectors is gaining wide-spread interest. A diamond is a wide-bandgap semiconductor with a bandgap of 5.47 eV and density of 3.52 g/cm$^{3}$. Owing to their exceptional physical properties, diamonds exhibit unique and superior performance in radiation and particle detection. In addition to outstanding radiation hardness, diamonds offer several advantages: low leakage current when used as a semiconductor detector~\cite{DiamondLeak}, the presence of color centers in the visible wavelength range for scintillator applications~\cite{DiamondOpticalCenter}, and the highest Debye temperature among crystals for bolometers~\cite{DiamondDebye}. The three types of diamond-based detectors have been used for radiation detection~\cite{diamond_semicon}~\cite{diamond_syn_sc1}~\cite{diamond_bolo}, and further advancements in their applications are anticipated. 
Among these, scintillators are particularly practical because they require less complex instrumentation, making them easier to operate and more accessible to general users through simplified data acquisition systems.
Current research on novel scintillators is largely focused on achieving higher light yield, faster response (e.g., shorter decay time constant), and tunable luminescence wavelengths. These goals can be achieved either by discovering new materials or by improving existing ones. In the case of synthetic diamond scintillators, we found that nitrogen impurities substituting for carbon atoms in the crystal act as color centers ~\cite{diamond_syn_sc1}. At present, synthetic diamonds can be stably manufactured using chemical vapor deposition and high-pressure high-temperature methods ~\cite{CVD}~\cite{HPHT}, with the impurity concentration controlled by intentionally adding or removing specific elements.
Because impurity defects other than nitrogen may also function as color centers, there is significant potential for further improving the luminescence characteristics. 
However, the mechanisms underlying radiation-induced luminescence in diamonds remain unclear. Therefore, it is essential to identify the key processes that govern enhancement in luminescence performance.

In this study, we present a time-dependent density functional theory (TDDFT) ~\cite{RungeGross1984} approach to investigate changes in the density of states (DOS) in diamonds, aiming to clarify the electron behavior following radiation excitation. A realistic diamond crystal model was constructed, with impurity dopants optionally introduced to evaluate features such as color centers relevant to diamond scintillators. 
The relationship between the DOS after alpha-ray irradiation and the presence of impurities was systematically analyzed.
Combined examination of electron and lattice dynamics under radiation excitation establishes a computational framework for elucidating the fundamental mechanisms of radiation detection in diamonds, with applicability extending beyond scintillators to other types of detectors. %Furthermore, it evaluates detector performance as a function of impurity content, with the aim of accelerating the development of diamond-based radiation detectors.

\section{Method}
To follow the excitation process in diamonds upon alpha-ray irradiation, molecular dynamics (MD) simulations were performed by treating a moving helium$^{2+}$ (He) ion with a kinetic energy of 600 keV or 5 MeV passing through a crystalline diamond. 
These kinetic energies were selected for the following reasons: 5 MeV corresponds to the typical alpha-ray energy from commercially available radiation sources, while 600 keV corresponds to the energy at which the maximum energy loss—commonly referred to as the energy deposition per unit length (dE/dx)— is expected.
Using SRIM simulations~\cite{SRIM_sim}, the dE/dX of helium in a diamond at kinetic energies of 600 keV and 5 MeV were calculated to be 679 keV/$\mu$m and 285 keV/$\mu$m, respectively.

\begin{figure}[htbp]
 \begin{center}
 \includegraphics[width=5.5cm]{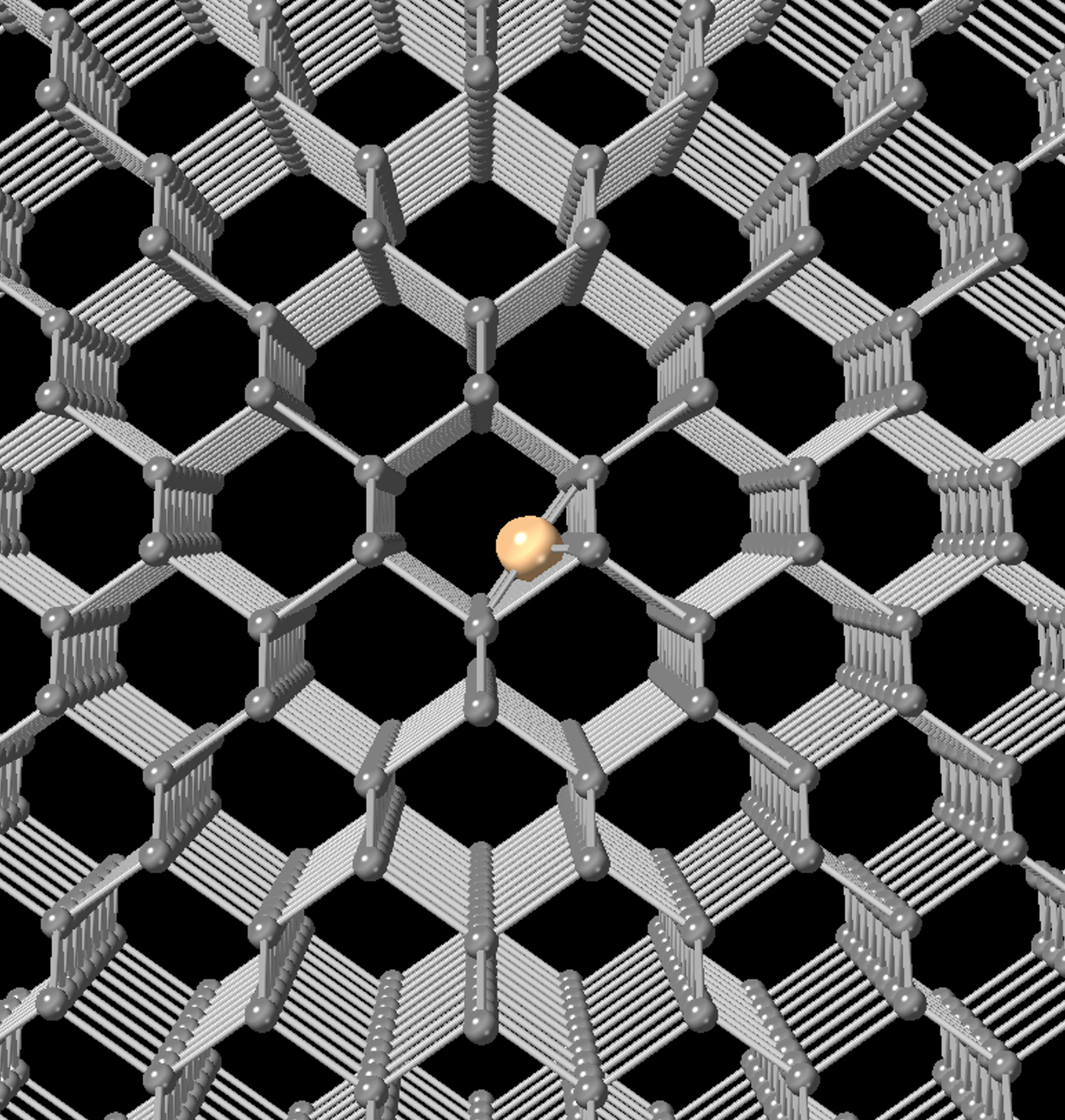}
\caption{Schematic image of the bulk diamond model with a He ion.}
\label{fig:diamond_models}
 \end{center}
\end{figure}

A diamond structural model was constructed by adopting the cubic unit cell of bulk crystal and by periodically replicating it to form a 3$\times$3$\times$3 supercell, yielding a periodic structure with a lattice dimension of approximately 1.08 nm.
This supercell contains 216 carbon atoms. A slab model, consisting of a few surface layers with vacuum elsewhere, was also examined as an alternative representation of the diamond structure. However, owing to structural collapse caused by Coulomb explosion following helium passage, the slab model was not adopted. By contrast, the bulk model remained stable without undergoing Coulomb explosion.
To simulate doping with a potential color center, a carbon atom near the center of the supercell was replaced with an impurity atom.
A schematic image of the bulk diamond model is shown in Fig.~\ref{fig:diamond_models}. 
Gray spheres represent carbon atoms, while the bright orange sphere represents a He ion.
When simulating the passage of the He ion, its initial position was arbitrarily set to avoid lattice collisions along the (001) direction, corresponding to the vertical direction in Fig.~\ref{fig:diamond_models}.
After passing through several times with a period of 1.08 nm along the (001) direction, evolving changes in the crystal structure and DOS were observed.
Increasing the number of passages enhanced the energy loss inside the crystal.

Coupled to the MD simulation, the time evolution of the electron wavefunction was computed within a TDDFT scheme. A plane waves basis set with a cutoff energy of 64 Ry was used to express the time-dependent Kohn-Sham orbitals. The many-body electron-electron interaction was treated as a density functional theory (DFT) scheme, in which the exchange-correlation potential was expressed as a functional of the electron charge density.\cite{Kohn1964, KohnSham1965} Here, the local density approximation (LDA) was employed using a functional form\cite{PerdewZunger} fitted to the numerical results of electron gas\cite{Ceperley-Alder}. Interactions between valence electrons and ions were represented by using norm-conserving pseudopotentials\cite{TM} with separable form in the nonlocal part\cite{KB}. The real-time evolution of electrons' wavefunctions within TDDFT was obtained by numerically solving the time-dependent Kohn-Sham equation
\begin{eqnarray}
i\frac{\partial }{\partial t}\psi^{TDDFT}_{\rm m}({\bf r},t)=H^{KS}({\bf r},t)\psi^{TDDFT}_{\rm m}({\bf r},t),
\label{TDKSE}
\end{eqnarray}
where the $\psi^{TDDFT}_{\rm m}({\bf r},t)$ represents the time-dependent Kohn-Sham orbital at the band index ${\rm m}$,  and $H^{KS}({\bf r},t)$ denotes the Kohn-Sham Hamiltonian as a function of space ${\bf r}$ and time $t$. 
For the numerical computation, the split-operator scheme\cite{Suzuki1992,Suzuki1993}, which is implemented for the plane wave scheme\cite{SuginoMiyamoto, SuginoMiyamoto2}, was used. Molecular dynamics simulation was performed for all ions according to the force field that was dynamically modulated upon He ion irradiation. All numerical computations employed a time step of 0.005 atomic units (= 0.121 attoseconds). An open-source code {\sc fpseid}$^{21}$\cite{FPSEID21} was used for computations, and the MD simulations were carried out under constant-volume and constant-total-energy (potential plus ionic kinetic energy) condition.

The time evolution of the DOS obtained by the TDDFT-MD simulation $DOS^{TDDFT}(\epsilon,t)$ at electron energy $\epsilon$ at $t$ is given by
\begin{eqnarray}
DOS^{TDDFT}(\epsilon,t)=\frac{1}{\pi}Im\sum_{{\rm m}.{\rm n}}
\frac{|<\psi^{TDDFT}_{\rm m}({\bf r},t)\mid \psi^{DFT}_{\rm n}({\bf r},t)>|^2}
{\epsilon_{\rm n}-\epsilon+io^+},
\label{DOS}
\end{eqnarray}
where $\psi^{DFT}_{\rm n}({\bf r},t)$ represents the Kohn-Sham orbital with band index ${\rm n}$, obtained by static DFT calculations using the atomic coordinates of a snapshot of the TDDFT-MD simulation at time $t$. 
The basic calculations were conducted assuming absolute zero temperature, such that at time zero the lattice (excluding He ions) is perfectly stationary ($T_{\mathrm {init}}$ = 0), as discussed in subsections 3.1 to 3.2. The effects of finite temperature are addressed in subsection 3.3.

\section{Result}

\subsection{Non-doped diamond}
\begin{figure}[htbp]
 \begin{center}
 \includegraphics[width=12cm]{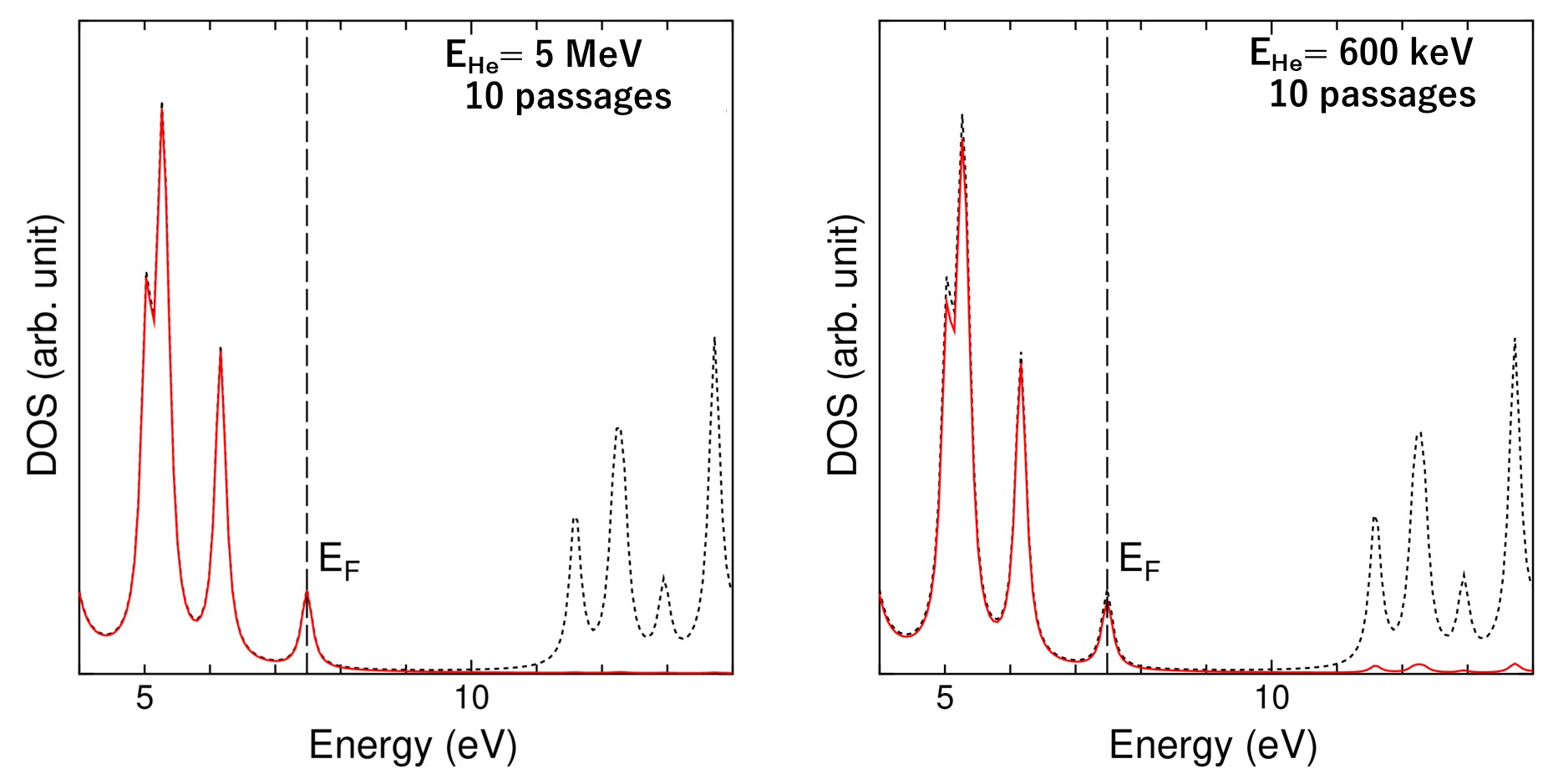}
\caption{DOS distribution after 10 passages of a He ion through a non-doped diamond crystal: 5 MeV He ion (left panel) and 600 keV He ion (right panel).}
\label{fig:DOS_E_comp}
 \end{center}
\end{figure}

Fig.~\ref{fig:DOS_E_comp} shows the DOS distributions for the non-doped diamond after 10 successive passages of the He ion.
The horizontal axis represents electron energy levels, while the vertical axis shows the DOS in arbitrary but consistent units.
The red solid line and black dotted line denote the TDDFT and ground-state DFT results, respectively.
The Fermi energy ($E_{F}$) position is indicated by a vertical black dashed line. 
For the 5 MeV He ion, the electron excitation into the conduction band was negligible. By contrast, for the 600 keV He ion, several high-energy peaks appeared, indicating significant electron excitation into the conduction band. As previously mentioned, this difference is attributed to variations in dE/dx between the two cases, suggesting that sufficient energy loss can induce electron excitation into the conduction band in non-doped diamonds.
Next, we focused on measurements using an alpha source and conduct simulations with 5 MeV He ions.

\subsection{n-type doping diamond}
\begin{figure}[htbp]
 \begin{center}
 \includegraphics[width=12cm]{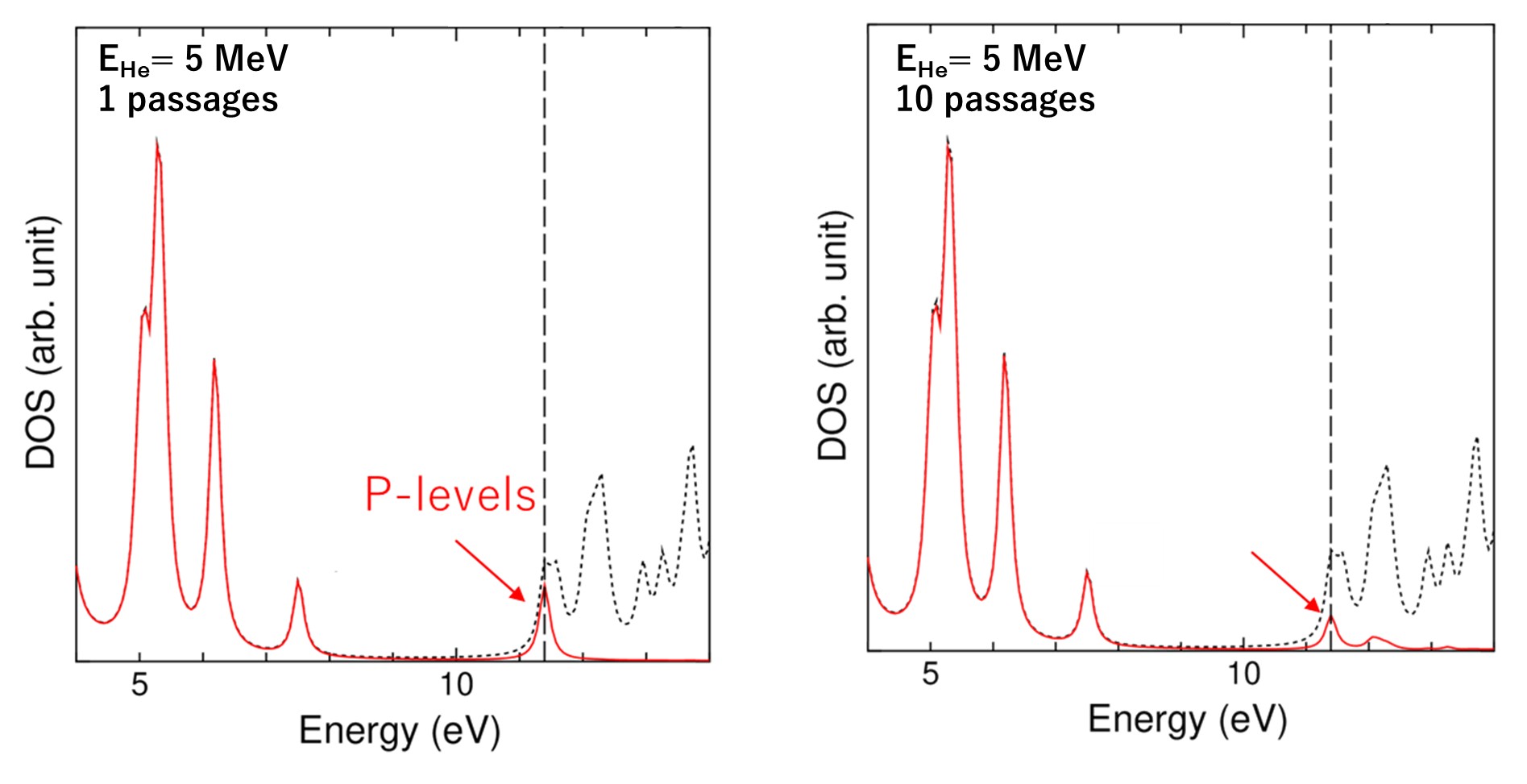}
\caption{DOS distribution after passing a 5 MeV He ion through a P-doped diamond crystal.}
\label{fig:DOS_E_comp2}
 \end{center}
\end{figure}

The DOS distribution was calculated for a phosphorus-doped (P-doped) diamond ~\cite{P-dopedCVD}.
Fig.~\ref{fig:DOS_E_comp2} presents a comparison of the DOS distributions after several 5 MeV He ion passages.
Comparison with Fig.~\ref{fig:DOS_E_comp} shows that P-doping shifted the Fermi level closer to the conduction band (denoted as P-level in Fig.~\ref{fig:DOS_E_comp}). 
As the number of He ion passages increased, the P-level shifted toward higher energies.
%Additionally, the DOS distributions after 10 passages reveals that P-doping significantly enhances electron excitation into the conduction band compared to non-doped diamond.
This behavior is consistent with the typical characteristics of an n-type doped semiconductor and supports the validity of the computational approach and diamond structural model employed in this study.

\begin{figure}[htbp]
 \begin{center}
 \includegraphics[width=10.5cm]{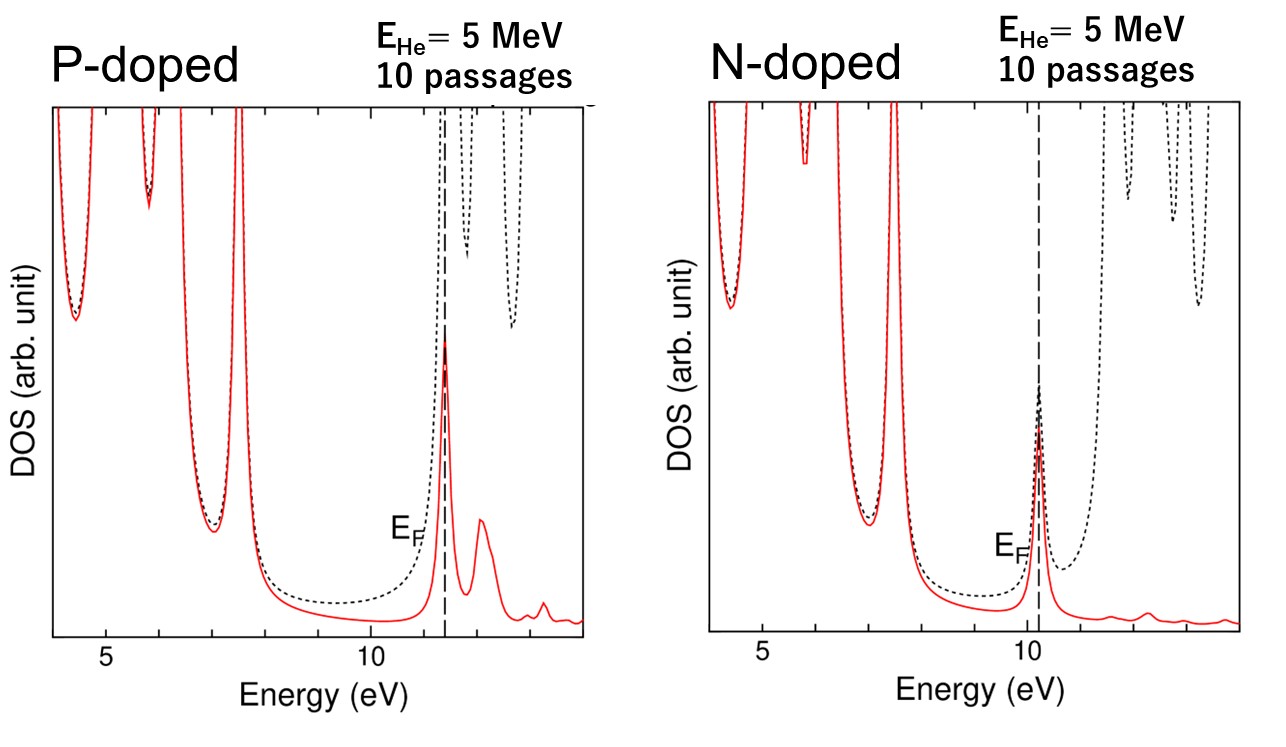}
\caption{DOS distribution after 10 passages of a 5 MeV He ion through P-doped (left panel) and N-doped (right panel) diamonds. Note that the unit of the vertical axes representing DOS are common in this Figure but 10 times smaller than that in Figs.~\ref{fig:DOS_E_comp},~\ref{fig:DOS_E_comp2},  and ~\ref{fig:DOS_temporal} (see below).}
\label{fig:DOS_E_comp3}
 \end{center}
\end{figure}

Subsequently, nitrogen-doped (N-doped) diamond was modeled, and the DOS distribution was calculated.
Fig.~\ref{fig:DOS_E_comp3} compares the DOS distributions of P-doped and N-doped diamonds, scaled around the Fermi level.
The results indicate that the N-level was deeper within the band gap than the P-level, which is theoretically consistent~\cite{DonorLevel}. %Due to its shallower donor level, P-doping more effectively promotes excitation to higher energy levels under the same energy-loss conditions, resulting in an increased DOS observed at higher energies.
By integrating the height of the DOS along the energy axis, the number of excited states (i.e., electrons) can be calculated; however, regardless of the dopant, the value was approximately 0.3 e. This discrepancy arises from the broadening factor applied during the DOS calculation, which introduces a significant error.
By increasing the number of statistical samples and continuing the calculations until the alpha particles stop, an evaluation with significance beyond the error is expected.

\begin{figure}[htbp]
 \begin{center}
 \includegraphics[width=13.5cm]{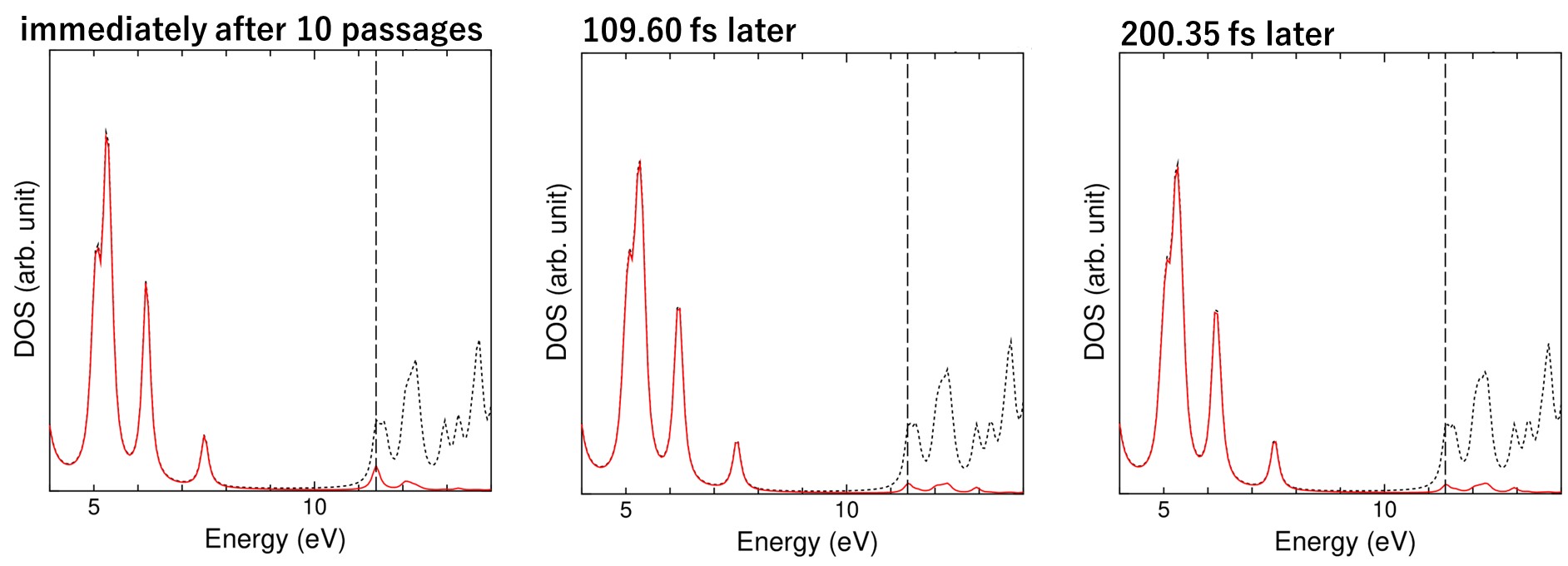}
\caption{Temporal changes in the DOS after 10 passages of a 5 MeV He ion through a P-doped diamond. }
\label{fig:DOS_temporal}
 \end{center}
\end{figure}

To further investigate the excited states in the P-doped diamond, the temporal evolution of the DOS following He ion passage was analyzed.
After 10 passages of the 5 MeV He ion, the ion was rapidly removed from the model, and time-dependent DOS calculations were performed up to approximately 200 fs. Although this rapid removal does not reflect natural dynamics, no significant numerical instabilities were observed during the time-evolution.
The results are shown in Fig.~\ref{fig:DOS_temporal}.
The DOS of the excited electrons exhibited no significant decrease, indicating that the P-level in the diamond remained in an excited state for more than 200 fs. Since 200 fs is a sufficiently long timescale to assess the stability of the excited states, the persistence of these elevated energies suggests that returning to the ground or stable state requires energy release through some process, which may be detectable as particle signals--analogous to luminescence in a scintillator.

\begin{figure}[htbp]
 \begin{center}
 \includegraphics[width=7cm]{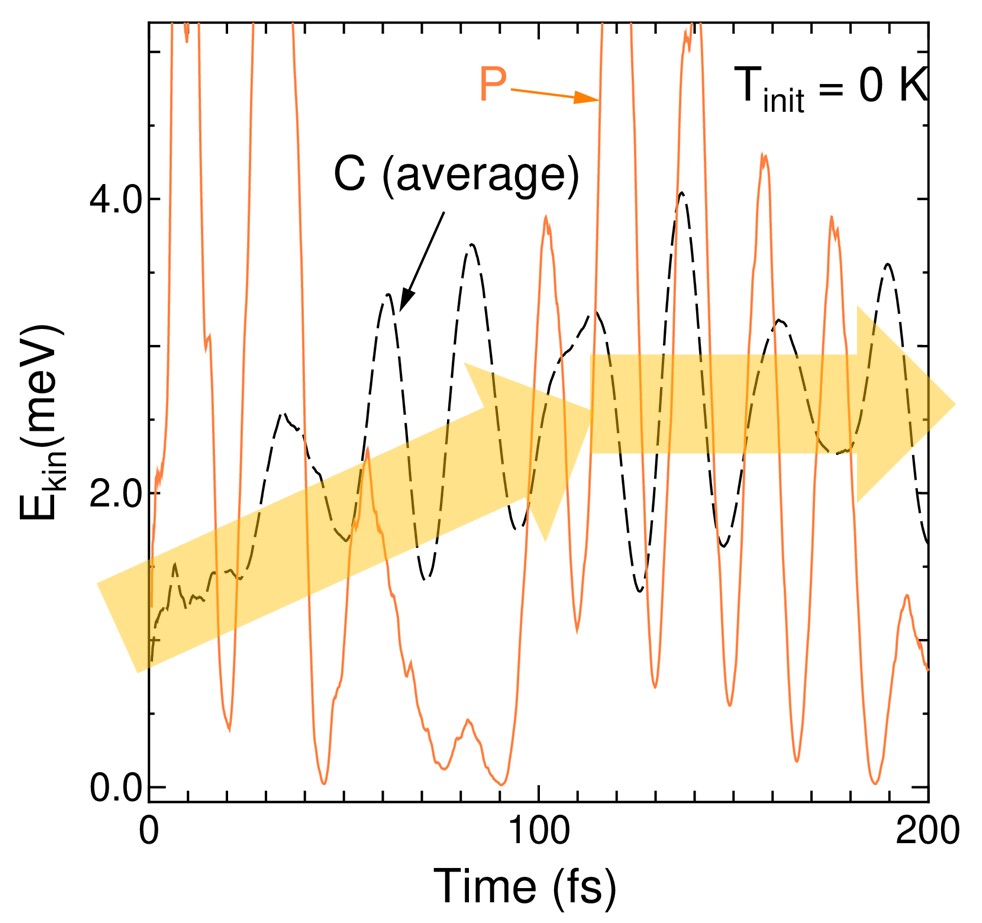}
\caption{Time-resolved lattice kinetic energies of carbon atoms and the P donor following the removal of the He ion after 10 passages with a kinetic energy of 5 MeV through the P-doped diamond. The yellow arrows indicate the trend in the time-evolution of the lattice kinetic energies.}
\label{fig:LatticeKinetic}
 \end{center}
\end{figure}

Fig.~\ref{fig:LatticeKinetic} shows the temporal evolution of lattice kinetic energies following the removal of the He ion.
The black dashed line indicates the average kinetic energy of carbon atoms in the diamond model, while the solid orange line denotes that of the P donor. The figure presents the time evolution up to 200 fs immediately after He ion removal.
The initially small kinetic energy gradually increased to several meV. Oscillations are observed in both curves, arising not only from fluctuations inherent in the MD calculations but also from the finite periodic structure of the diamond model.
Around 100 fs, the kinetic energy of the carbon atoms appeared to stabilize, as indicated by the yellow arrow in the figure.
For the P donor, fluctuations remained large even at 200 fs, suggesting that calculations over a longer timescale are required to determine whether the system reaches thermal equilibrium.
%Not only the energy of the excited electrons but also the kinetic energies of the carbon atoms and the P-donor remain elevated beyond 200 fs. 

\subsection{Temperature dependence}
\begin{figure}[htbp]
 \begin{center}
 \includegraphics[width=14cm]{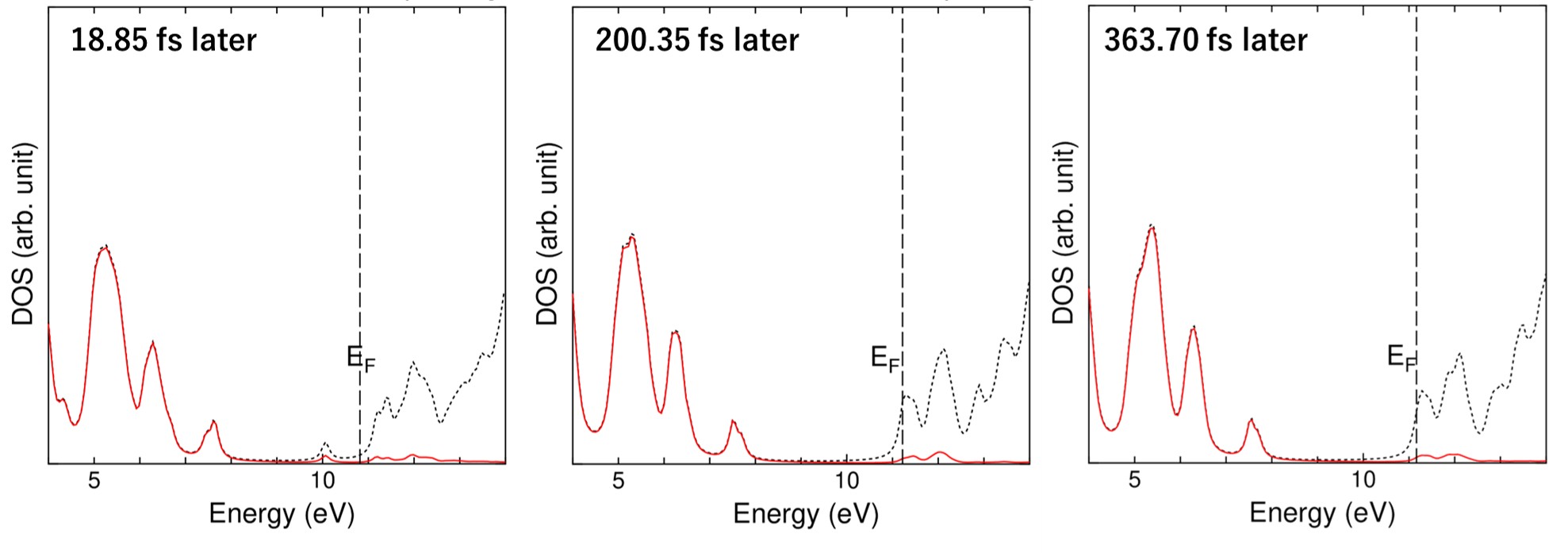}
\caption{Temporal changes in the DOS following the removal of the He ion after 10 passages with a kinetic energy of 5 MeV through the P-doped diamond with a lattice temperature of 300 K.}
\label{fig:DOS_RT}
 \end{center}
\end{figure}

The temporal changes in the DOS distribution and temporal evolution of lattice kinetic energies were also calculated at room temperature. The time-averaged temperature was set to approximately 300 K, denoted as ($T_{\mathrm {avr}}$ = 300 K). Fig.~\ref{fig:DOS_RT} shows the DOS distributions at 18.85 fs, 200.35 fs, and 363.70 fs (from left to right), following the removal of the He ion after 10 passages with a kinetic energy of 5 MeV.
A shift was observed in the DOS associated with excited states over time, as seen when comparing the results at 18.85 fs and 200.35 fs.
This behavior may be attributed to shifts in the DFT reference level (indicated by the black line) due to changes in atomic coordinates at each snapshot, suggesting a possible coupling.
Because the results at 200.35 fs and 363.70 fs exhibited minimal difference, the system likely approached equilibrium around 200 fs.
These results also indicate that electrons may remain in excited states for a timescale exceeding 300 fs.

\begin{figure}[htbp]
 \begin{center}
 \includegraphics[width=7.5cm]{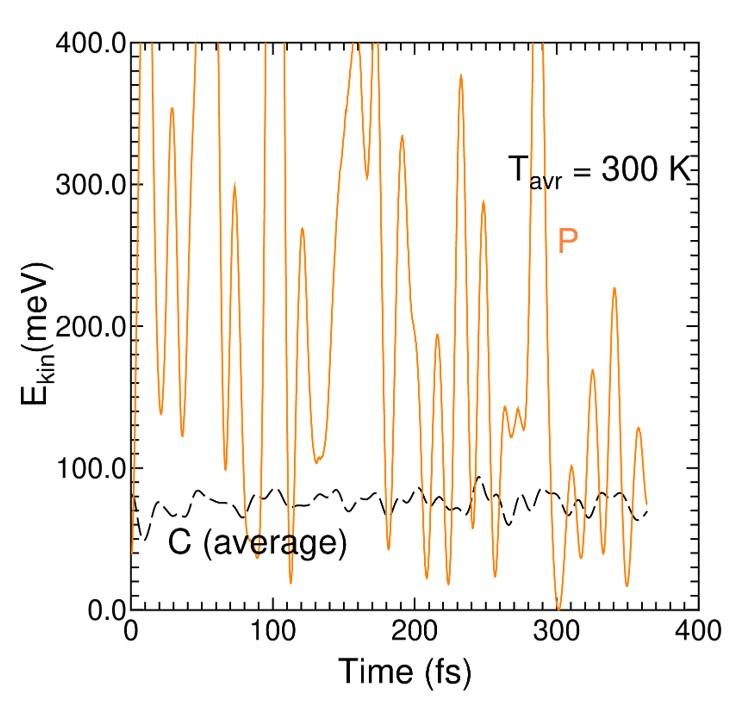}
\caption{Time-resolved lattice kinetic energies of carbon atoms and the P donor following the removal of the He ion after 10 passages with a kinetic energy of 5 MeV through the P-doped diamond with a lattice temperature of 300 K.}
\label{fig:LatticeKinetic300}
 \end{center}
\end{figure}

Fig.~\ref{fig:LatticeKinetic300} shows the time-resolved lattice kinetic energies of carbon atoms and the P donor after the removal of the He ion.
At 300 K, the kinetic energy increased by several tens of times compared with its value at 0 K, reflecting the thermal energy of each atom. The lattice underwent thermal vibrations, which lead to larger temporal fluctuations in the kinetic energy, especially for the P donor, owing to the presence of only one P atom in the unit cell.
The P donor reached convergence around 300 fs, whereas for carbon, no clear sign of energy dissipation was observed even beyond 300 fs. These results further suggest the necessity of a mechanism for energy release.

\newpage

\section{Discussion}
The developed computational method produced results consistent with qualitative estimates, including the effects of donor levels and temperature, and demonstrated that the DOS distribution undergoes significant changes upon irradiation with He ions passing through a diamond unit cell.
The present model revealed differences in the DOS distribution; however, the dopant did not exhibit any statistically significant effect on the number of excited electrons. 
To clarify the effects of doping, either the current computational method must be improved to better accumulate the energy loss from alpha-rays or calculations involving high-energy nuclear events with larger dE/dx values must be performed.
The DOS of the excited electrons did not show a significant decrease within 200 fs. 
If impurity-induced defects act as color centers, a scintillation signal could be generated when excited electrons relax into those levels. In semiconductor detectors, by contrast, the excited electrons serve as carriers that are extracted as electrical signals.
Continued advancement of the computational method is essential to enable its application to the experimental development of diamond-based detectors in the near future. In particular, investigating impurity concentrations, including donors, acceptors (e.g., boron), or combinations of both, will be crucial for a comprehensive evaluation of diamond detectors.

\section{Conclusion}
We proposed a computational approach based on real-time time-dependent density functional theory to assess the performance of diamond-based radiation detectors.
Diamond structural models were constructed using a 3$\times$3$\times$3 supercell of a cubic bulk crystal unit cell, and the temporal evolution of the electronic density of states and lattice kinetic energies was calculated following the traversal of a He ion simulating an alpha-ray. 
The method also accommodates impurity doping, enabling analysis of both the DOS distribution and the time-resolved lattice kinetic energies of atoms.
Time-evolution calculations following the removal of the He ion showed that the excitation persists for more than 200 fs, suggesting a potential energy relaxation pathway detectable as a signal in particle detectors.

\section*{Acknowledgement}
The calculations were performed using the NEC SX-Aurora TSUBASA of the Earth Simulator (ES4) in Japan. This work was conducted as part of the Challenge Usage Project of Earth Simulator in FY2024 and as a Proposed Research Project in FY2025 (1-25021).
This study was supported by the Japan Society for the Promotion of Science (JSPS) KAKENHI Grant-in-Aid for Early-Career Scientists 24K17075.

%%%\bibliography{mybibfile}

\bibliographystyle{elsarticle-num} 

\bibliography{UmemotoMiyamoto}

\end{document}